\documentstyle[prl,aps,epsf]{revtex} \def\narrowtext{} \tighten \twocolumn
\input epsf.sty 
\begin{document}
 
\title{COMMENT ON RECENT TUNNELING MEASUREMENTS ON 
	Bi$_{2}$Sr$_{2}$CaCu$_{2}$O$_{8}$}
\author{
        Z. Yusof, L. Coffey
       }
\address{
         Illinois Institute of Technology, Chicago Illinois 60616
        }


\maketitle
\narrowtext

Recent PCT and STM measurements of the tunneling conductance on 
Bi$_{2}$Sr$_{2}$CaCu$_{2}$O$_{8}$ (Bi-2212)\cite{Yannick}\cite{Renner}
with various hole doping
reveal an interesting asymmetry in the conductance peaks.
In all cases, the negative bias peak is higher than the positive bias peak. 
The purpose of this comment is to point out the potential significance
of this type of conductance peak asymmetry in determining the symmetry
of the underlying superconducting order parameter. 

\begin{figure}
\vspace{0.0cm}
\epsfysize=5.0in
\epsfbox{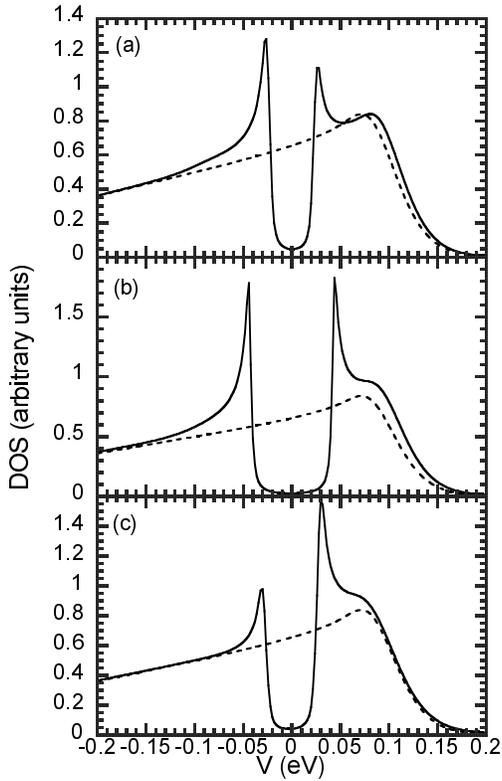}
\vspace{-1.7cm}
\caption{Tunneling DOS for (a) $d_{x^{2}-y^{2}}$ gap, (b) s-wave gap, and
(c) $s_{x^{2}+y^{2}}$ gap. All DOS were obtained using $\Delta_{o}=0.043$ eV,
$\gamma=0.002$ eV, $\theta=0.25$ rad, $\theta _{o}=0.05$ rad. $c_{o}$
in the band structure has been modified to 0.2 eV to minimize the influence
of the upper band edge. The dashed lines correspond to the normal state.
Refer to Ref. 3 for definitions of the parameters.}
\label{fig1}
\end{figure}

A calculation by the 
present authors\cite{Yusof} of the tunneling density of states (DOS) for
$d_{x^{2}-y^{2}}$ superconductors incorporating group velocity, tunneling
directionality,
and a realistic band structure extracted from ARPES indicates that such peak
asymmetry seen in tunneling experiments may be an intrinsic property of
$d_{x^{2}-y^{2}}$ pairing. 

In this comment, we explore this issue further by comparing calculations of
the tunneling DOS for $d_{x^{2}-y^{2}}$
($\Delta ({\bf k})=\Delta _{o}[\cos (k_{x}a)-\cos (k_{y}a)]/2$),
isotropic s-wave ($\Delta=\Delta _{o}$), and $s_{x^{2}+y^{2}}$
($\Delta ({\bf k})=\Delta _{o}[\cos (k_{x}a)+\cos (k_{y}a)]$)
gap symmetries.
Figure 1 displays typical results for the three gap symmetries.
Figure 1(a) for the $d_{x^{2}-y^{2}}$ state illustrates
that the
negative energy peak is higher than the positive energy peak, whereas in
Fig. 1(c), the opposite is the case for the $s_{x^{2}+y^{2}}$ symmetry.
This peak asymmetry remains the same with variations in chemical potential
and tunneling direction in the model and appears to be an intrinsic
and robust property of
directional tunneling into a $d_{x^{2}-y^{2}}$ or $s_{x^{2}+y^{2}}$ state.
Negative values of V in Fig. 1 correspond to electron extraction from the
superconductor while positive values of V correspond to electron
injection into the superconductor. This sign convention is the same as
in the experiments.\cite{Yannick}\cite{Renner}  
The isotropic s-wave DOS (Fig. 1(b)) shows
approximately symmetric peaks, with the slight asymmetry here tending
to reflect the shape of the underlying normal state DOS.
These results suggest that the peak asymmetry in the recent optimal
Bi-2212 data\cite{Yannick}\cite{Renner}
may be a subtle signature of a $d_{x^{2}-y^{2}}$ gap. 

Theoretical studies of high-T$_{c}$ superconductors indicate that a change in
symmetry, from $d_{x^{2}-y^{2}}$ to
$s_{x^{2}+y^{2}}$, may occur in the overdoped region of the
phase diagram.\cite{Koltenbah} Figure 1 illustrates that such a change
in the symmetry would be accompanied
by a switching of the asymmetry of the conductance peaks in tunneling
conductance measurements.
However, the most likely example of cuprate oxide s-wave superconductivity
appears to be the electron doped NdCeCuO\cite{Anlage} which makes it
tempting to speculate that one
might anticipate a change in gap symmetry from $d_{x^{2}-y^{2}}$ to
$s_{x^{2}+y^{2}}$
to occur towards the underdoped region of the phase diagram
instead.

We acknowledge conversations with J.F. Zasadzinski and N. Miyakawa.

\end{document}